\documentclass{achemso}
\setkeys{acs}{doi = true}

\usepackage[version=3]{mhchem} 
\usepackage{xcolor}
\usepackage{soul}
\usepackage{chemfig}
\usepackage{hyperref}
\newcommand{\doi}[1]{\href{http://dx.doi.org/#1}{\nolinkurl{#1}}}
\usepackage{comment}
\usepackage{amssymb}

\author{Ana M. Valencia}
\affiliation{Carl-von-Ossietzky Universit{\"a}t Oldenburg, Institute of Physics, Germany}
\email{ana.valencia@uni-oldenburg.de}
\author{Lisa Schraut-May}
\affiliation{Julius-Maximilians-Universit\"at, W\"urzburg, Experimental Physics VI, Germany}
\author{Marie Siegert}
\affiliation{Julius-Maximilians-Universit\"at, W\"urzburg, Experimental Physics VI, Germany}
\author{Sebastian Hammer}
\affiliation{Julius-Maximilians-Universit\"at, W\"urzburg, Experimental Physics VI, Germany}
\author{Beatrice Cula}
\affiliation{Humboldt-Universit\"at zu Berlin, Department of Chemistry, Germany}
\author{Alexandra Friedrich}
\affiliation{Julius-Maximilians-Universit\"at W\"urzburg, Institute of Inorganic Chemistry \& Institute for Sustainable Chemistry and Catalysis with Boron (ICB), Germany}
\author{Holger Helten}
\affiliation{Julius-Maximilians-Universit\"at W\"urzburg, Institute of Inorganic Chemistry \& Institute for Sustainable Chemistry and Catalysis with Boron (ICB), Germany}
\author{Jens Pflaum}
\affiliation{Julius-Maximilians-Universit\"at, W\"urzburg, Experimental Physics VI, Germany}
\altaffiliation{Center for Applied Energy Research e.V. (CAE Bayern), W\"urzburg, Germany}
\author{Caterina Cocchi}
\affiliation{Carl-von-Ossietzky Universit{\"a}t Oldenburg, Institute of Physics, Germany}
\altaffiliation{Friedrich-Schiller-Universit\"at Jena, Institute of Condensed Matter Theory and Optics, Germany}
\email{caterina.cocchi@uni-jena.de}
\author{Andreas Opitz}
\affiliation{Humboldt-Universit\"at zu Berlin, Institut f\"ur Physik, Germany}
\email{andreas.opitz@hu-berlin.de}

\title{Unveiling the Role of Solvents in DBTTF:HATCN Ternary Cocrystals}

\begin{document}

\begin{tocentry}
\centering
\includegraphics[width=6.8cm]{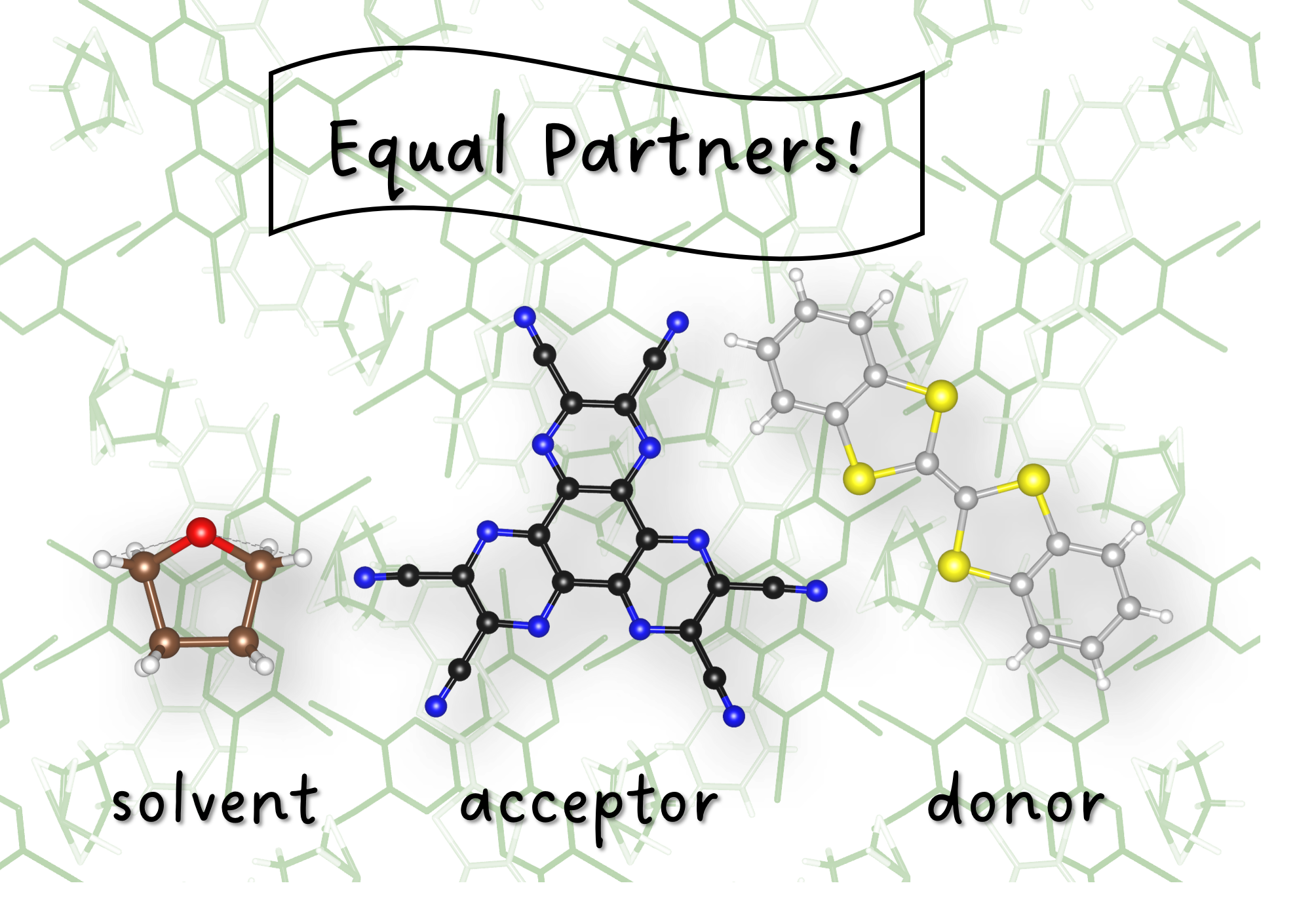}

\end{tocentry}

\newpage
\begin{abstract}
Donor-acceptor (D:A) cocrystals offer a promising platform for next-generation optoelectronic applications, but the impact of residual solvent molecules on their properties remains an open question. We investigate six novel D:A cocrystals of dibenzotetrathiafulvalene (DBTTF) and 1,4,5,8,9,11-hexaazatriphenylenehexacarbo-nitrile (HATCN), prepared via solvent evaporation, yielding 1:1 molar ratios, and horizontal vapor deposition, resulting in solvent-free 3:2 cocrystals. Combining spectroscopy and density-functional theory (DFT) calculations, we find that, while the electronic and optical properties of the cocrystals are largely unaffected by solvent inclusion, the charge transfer mechanism is surprisingly complex. Raman spectroscopy reveals a consistent charge transfer of 0.11~$e$ across all considered structures, corroborated by DFT calculations on solvent-free systems. Partial charge analysis reveals that in solvated cocrystals, solvent molecules actively participate in the charge transfer process as primary electron acceptors. This involvement can perturb the expected D:A behavior, revealing a faceted charge-transfer mechanism in HATCN  even beyond the established involvement of its cyano group. Overall, our study demonstrates that while solution-based methods preserve the intrinsic D:A characteristics, solvents can be leveraged as active electronic components, opening new avenues for material design.
\end{abstract}

\newpage


\section{Introduction}

Organic cocrystals (OCCs) are promising materials with tunable electronic and optical properties according to the selected molecular constituents and intermolecular interactions among them~\cite{gier+21aom,liu+23cec}. A key characteristic of OCCs, particularly those formed by donor and acceptor molecules, is the presence of a charge-transfer (CT) band in the absorption spectrum, resulting from electronic transitions involving primarily states localized on donor (D) or acceptor (A) molecules.\cite{Salzmann2015,mend+15ncom,cris+20jacs,liu-23cec,opit+22ma,goet+22mh,ourg+23jpcc} The properties of OCCs are highly sensitive to their crystal packing \cite{salz+16cgd,Wang+6jpcl2019} and donor:acceptor ratio~\cite{zhu+15small,salz+16cgd,suzu+20cgd,opit+22ma}. These structural and compositional variations directly influence the crystalline architecture, impacting the electronic characteristics of the material, including the degree of charge transfer (DCT).

Beyond intrinsic molecular design and stoichiometry, the presence of solvent molecules during crystallization can significantly influence the structural, electronic, and charge transport properties of OCCs.\cite{salz+16cgd,Wang+6jpcl2019,Zhai+21NC,zhu+15small} 
Residual solvents act as additional components in the cocrystals, interacting non-covalently with the donor and/or acceptor moieties and often altering the intermolecular pattern. For instance, recent studies have shown that solvent molecules can increase the donor-acceptor distances, weakening charge-transfer interactions and leading to changes in the structural and spectral characteristics~\cite{sun+20mcf,liu-23cec}. This discrete and yet transformative effect on the structural parameters, including stacking distances and slip-stacking angles, directly influences the electronic properties by modifying, for example, orbital overlap. As such, it provides a unique approach for fine-tuning material properties and charge-transfer interactions. Despite these intriguing observations, the precise role of specific solvents in templating cocrystal growth and their impact on the resulting charge-transfer properties remains an unexplored area that requires further detailed investigation.

In this study, we investigate the influence of residual solvent molecules in OCCs formed by the established donor molecule dibenzotetrathiafulvalene (DBTTF)\cite{B810993A} and the strong electron acceptor 1,4,5,8,9,11-hexaazatriphenylenehexacarbo\-nitrile (HATCN).\cite{kana+86joc,broe+10prl,Galle+09sm,chif+10} 
The electron-donating ability of DBTTF is primarily attributed to its electron-rich tetrathiafulvalene core and extended $\pi$-conjugation from fused benzene rings. This molecular architecture facilitates a low oxidation potential and the formation of stable radical cation and dication species.\cite{geng+14caej,li+11pccp,baye+19cm} On the other hand, the high electron affinity of HATCN, with experimental values ranging from 5.7 to 6.1 eV,\cite{kim+09apl,lee+12oe} is attributed to its six electron-withdrawing nitrile groups and its deep lowest unoccupied molecular orbital (LUMO) energy. We report the formation of six novel DBTTF:HATCN cocrystals, prepared using horizontal vapor deposition (HVD) and solvent evaporation with three different solvents: tetrahydrofuran (THF), acetone (ACE), and acetonitrile (MeCN). We investigate their electronic and optical properties both in solvent-free ($n=0$) and solvated forms ($n=1, 2$, where $n$ is the number of solvent molecules per D:A pair). These OCCs are prepared via solvent evaporation, yielding five cocrystals with a 1:1 donor:acceptor ratio (D:A:sol$_\mathrm{n}$), or through horizontal vapor deposition to obtain solvent-free samples with a 3:2 donor:acceptor ratio (D$_3$:A$_2$). We characterize the optical properties of these systems using absorbance spectroscopy, while vibrational analysis based on Raman spectroscopy is employed to determine the DCT~\cite{chap81jacs,kamp-neil89rissianchemrev}. Density-functional theory (DFT) calculations complement the experimental analysis, providing additional insight into charge distribution, electronic structure -- including the nature and spatial distribution of the frontier states -- and optical properties. This elaborated analysis offers a detailed understanding of the structure-property relationships in DBTTF:HATCN cocrystals, providing a robust rationale for the design, synthesis, and characterization of ternary co-crystals, where solvent molecules can be used as a versatile knob to customize material properties, tuning the interaction with and within the D:A pair.

\subsection{Structural Properties}

\begin{figure}
    \centering
\includegraphics[width=1.0\textwidth]{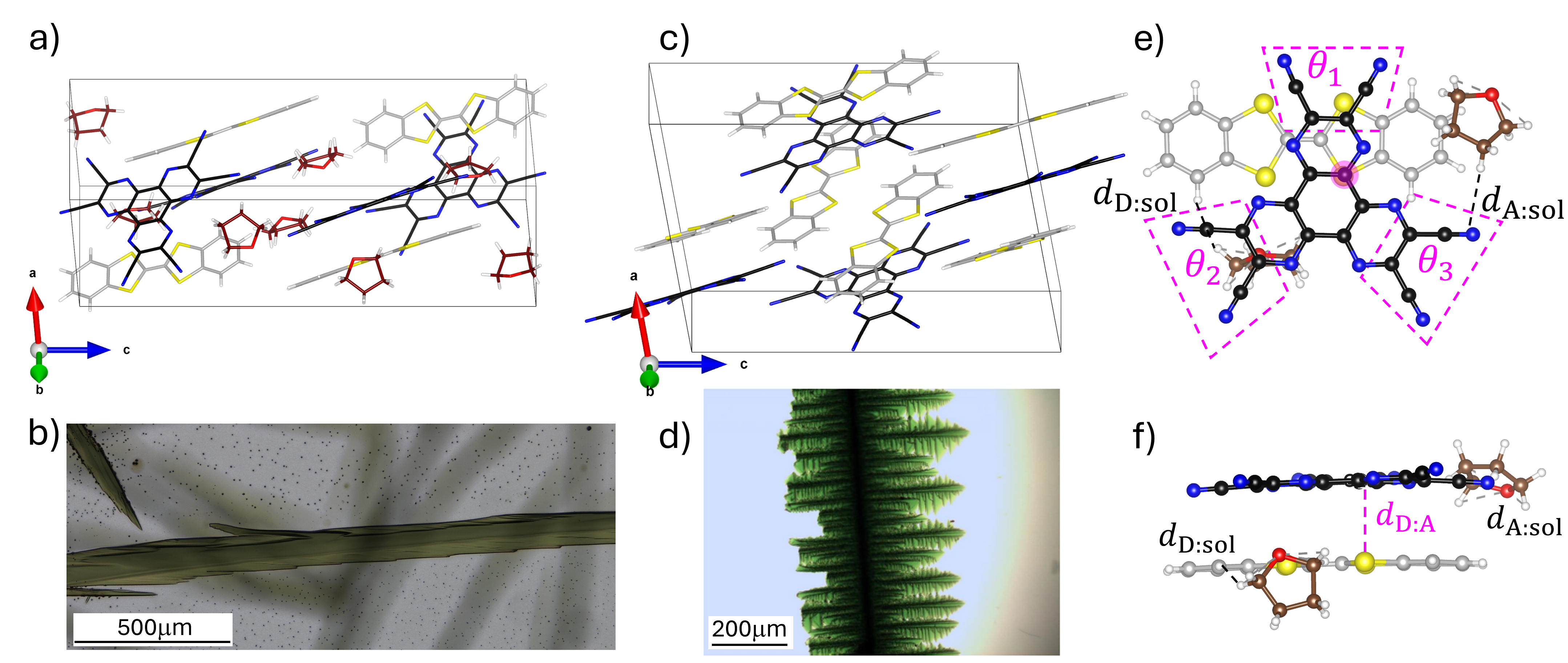}
    \caption{Unit cell representations, optical images, and structural details of DBTTF:HATCN:sol$_\mathrm{n}$ cocrystals. a), c) Stick representations of the unit cells of DBTTF:HATCN:THF$_2$ and DBTTF$_3$:HATCN$_2$, respectively, with their corresponding optical images in panels b) and d). e) Top and f) side views of the DBTTF:HATCN complex in the geometry extracted from the DBTTF:HATCN:THF$_2$ cocrystal, with $d_{D:A}$ being the shortest vertical distance between donor and acceptor, and  $d_{D:sol}$ ($d_{A:sol}$) the shortest distance between donor (acceptor) and solvent molecule. The dihedral angles along N$\equiv$C-C=C-C$\equiv $N, labeled $\theta_1$, $\theta_2$, $\theta_3$, are highlighted in magenta in panel e). S atoms are depicted in yellow, N atoms in blue, and H atoms in white. C atoms are color-coded by moiety: gray in DBTTF, black in HATCN, and brown in THF. O atoms are depicted in red. The structural drawings in panels a), c), e), and f) are generated with the visualization software VESTA~\cite{Momma-db5098}.}
    \label{fig:scheme}
\end{figure}

We successfully synthesized six stable cocrystalline polymorphs of DBTTF:HATCN, utilizing two distinct growth methods: solvent evaporation and HVD (see Experimental Section). All six OCCs exhibit a monoclinic crystal structure with the space group $P2_1/c$ (Figure \ref{fig:scheme}a and Table~\ref{tab:lattice-parameter}), exhibiting alternating D:A stacks. The five polymorphs grown through solvent evaporation consistently display a 1:1 D:A ratio and incorporate solvent molecules within their unit cells, forming needle-like crystals (Figure \ref{fig:scheme}b). These five solvated polymorphs share a common $\pi$-stacking motif, with solvent molecules interleaved between the D:A units. Although direct solvent-D:A interactions seem weak, their presence significantly influences the resulting crystal structure. Subtle variations occur in the lateral shifts of the molecules along the crystallographic direction $a$, leading to different intermolecular couplings. Nonetheless, the fundamental $\pi$-stacking arrangement remains consistent across these solvated polymorphs, suggesting that the interstitial spaces occupied by solvent molecules can be viewed as voids within the core stacking architecture.

Further analysis of solution-grown crystals reveals a consistent unidirectional stacking of HATCN and DBTTF molecules, indicating that strong and directional intermolecular interactions govern crystallization. Isomorphism~\cite{Haget2021} is observed for specific solvated crystals: for instance, ACE- and THF-containing OCCs with the same molar ratios are isostructural. This suggests that, at specific ratios, residual solvents do not fundamentally alter the underlying crystal packing. A common structural feature of these solution-grown OCCs is a reduced degree of molecular planarity compared to other related systems, which can influence intermolecular interactions and stacking arrangements. In these polymorphs, neighboring D:A stacks are arranged with herringbone packing.

In contrast to the solvated 1:1 OCCs, the polymorph grown by HVD exhibits a distinct 3:2 D:A ratio and forms dendritic crystals,  indicative of diffusion-limited aggregation (Figure~\ref{fig:scheme}c,d). Its crystal structure reveals D:A stacks where one and two alternating DBTTF molecules are positioned between HATCN, with neighboring HATCN molecules adopting a mirrored orientation. This 3:2 ratio in HVD-grown OCCs suggests a preferential accommodation of multiple donor molecules between acceptors in the absence of solvents. A DBTTF-saturated cocrystal with a 2:1 stoichiometry would be energetically unfavorable under these growth conditions. 

The HVD-grown cocrystals (3:2 ratio) form a brickwall arrangement between neighbouring stacks with two distinct stacking motifs characterized by alternating directions of HATCN molecules within the crystal lattice. This peculiar structure suggests the presence of different intermolecular distances compared to the uniformly oriented HATCN in solution-grown crystals. Likewise, the stacking of DBTTF molecules differs significantly between these two motifs, displaying alternating numbers and orientations, leading to a more complex arrangement in these structures. Overall, these different packing motifs result in varying magnitudes of $\pi$-stacking interactions, directly influenced by specific molecular arrangements and orientations. Notably, molecules in the HVD-grown crystals generally display a higher degree of planarity compared to their solution-grown counterparts. This characteristic could significantly impact $\pi$-orbital overlap and, consequently, electronic and optical properties.

The pronounced difference in the stoichiometry and packing motifs between solution-grown (1:1 ratio and herringbone arrangement) and HVD-grown OCCs (3:2, brickwall packing) highlights the significant influence of the local crystallization environment. Very likely, the relative sizes of the D and A molecules also play a role, as solution- and vapor-phase growth of D:A cocrystals usually proceed through different kinetic pathways, with solution growth potentially leading to kinetically limited phases where solvent molecules may not be able to completely diffuse out of the growing crystal lattice~\cite{ghos+18cgd}. The presence of solvent appears crucial for achieving the 1:1 stoichiometry, suggesting that interactions between donor and/or acceptor molecule and the solvent play a key role. Conversely, the absence of solvent during HVD growth seems to trigger a greater diversity of interaction pathways between the differently sized D and A molecules, resulting in the observed 3:2 ratio. Again, we cannot exclude the role of growth kinetics in the process~\cite{bric67jcg,dave+86jcg,ghos+18cgd}, but we will not focus on this aspect, since ultimately only the resulting packing and stoichiometry are relevant for the physical properties addressed in the following. Overall, these distinct packing motifs result in varying strengths of $\pi$-stacking interactions, directly influenced by the specific molecular arrangements and orientations.

\begin{table}
\centering
\caption{Structural properties of the investigated (solvated) DBTTF:HATCN cocrystals with the number of solvent molecules in the unit cell indicated by a subscript.  $Z_\mathrm{D:A}$ indicates the total number of donor and acceptor molecules, namely 4 each in the OCC with 1:1 ratio, as well as 6 DBTTF and 4 HATCN in the one with 3:2 ratio (D$_3$:A$_2$). The lattice parameters (lattice vectors $a$, $b$, $c$, and the angle $\beta$ enclosed between $a$ and $c$) are measured at 100~K. The distances between the donor and acceptor molecule ($d_{D:A}$), DBTTF and the solvent ($d_{D:sol}$), HATCN and the solvent ($d_{A:sol}$), and the dihedral angles $\theta_i$ (with $i=1,2,3$) along N$\equiv$C-C=C-C$\equiv $N in HATCN are evaluated for the experimental crystal structures.}
\label{tab:lattice-parameter}
\begin{tabular}{lcccccc}
\hline \hline
                      &D:A:ACE$_1$ &D:A:THF$_1$&D:A:MeCN$_1$ &D:A:ACE$_2$ & D:A:THF$_2$ & D$_3$:A$_2$  \\ \hline \hline
    $Z_\mathrm{D:A}$& 8 &8  & 8 &8  &8 & 10 \\\hline  \hline
    Volume [\AA{}$^3$] & 3351.67 & 3371.98 & 3221.69 & 3671.57 & 3713.75 & 3560.22 \\ \hline
    $a$ [\AA{}] &7.99&8.07 &8.23&7.72&7.70& 14.02  \\\hline                   
    $b$ [\AA{}] &16.62& 16.61&16.23&16.41&16.63& 12.07 \\ \hline
    $c$ [\AA{}] &25.52&25.29 &24.42&29.03 &29.10& 21.37     \\\hline
    $\beta$ [$^{\circ}$] &98.5& 95.9&99.0 &93.3&94.7& 100.1 \\\hline \hline
    $d_{D:A}$ [\AA{}] &3.28& 3.30&3.22 &3.27 &3.28  &   3.21; 3.34\\\hline
    $d_{D:sol}$ [\AA{}] &2.70& 2.79&2.89 &2.35 &2.54  &   -\\\hline
    $d_{A:sol}$ [\AA{}] &2.81& 2.47&2.94 &2.61 &2.77  &   -\\\hline \hline
    $\theta_1$ [$^{\circ}$] & 0.9 & 0.4 & 2.0 & 1.2 & 1.4 & 3.2 \\ \hline
    $\theta_2$ [$^{\circ}$] & 1.7 & 0.5 & 2.2 & 1.2 & 2.4 & 3.3 \\ \hline
    $\theta_3$ [$^{\circ}$] & 6.6 & 8.0 & 7.0 & 6.2 & 5.5 & 4.9 \\\hline \hline         
\end{tabular}
\end{table}


To understand the influence of residual solvent molecules on the structural properties of DBTTF: HATCN cocrystals, we performed DFT calculations starting from the experimentally resolved crystal structures (see Computational Details section). For each polymorph, we evaluated the shortest distance between D and A molecules along the $\pi$-stacking direction ($d_{D:A}$), as well as between the solvent and DBTTF ($d_{D:sol}$) and HATCN ($d_{A:sol}$), see Figure~\ref{fig:scheme}f. These intermolecular spacings give insights into the interaction strengths between the cocrystal components. In the D$_3$:A$_2$ OCCs, there are two $d_{D:A}$ values of 3.21~\AA{} and 3.34~\AA{} (Table~\ref{tab:lattice-parameter}), representing the distance between HATCN and each neighboring DBTTF rotated with respect to its sibling by almost 90$^{\circ}$ (see Figure~S2). This variation suggests two distinct D:A interactions in this binary cocrystal. In the solvated OCCs, $d_{D:A}$ ranges from 3.22~\AA{} to 3.30~\AA{} depending on the solvent type and concentration. However, the trend exhibited by $d_{D:A}$ is not consistent with the lattice parameter $a$, which is roughly aligned with the $\pi$-stacking direction of the D:A pairs. For example, the D:A:MeCN cocrystal has the largest value of $a$ and the smallest $d_{D:A}$ across all considered OCCs, see Table~\ref{tab:lattice-parameter}.  

This apparent contradiction can be solved by considering the interactions between the D:A pair and the solvent. The D:A:THF$_1$ cocrystal, exhibiting the largest $d_{D:A}$ among the considered OCCs, is characterized by the shortest distance between the acceptor and the solvent ($d_{A:sol}=2.47$~\AA{}). Conversely, in the D:A:MeCN cocrystal, the solvent molecules are the most distant from both the donor ($d_{D:sol}=2.89$~\AA{}) and the acceptor ($d_{A:sol}=2.94$~\AA{}), which are instead the closest to each other among the solvated OCCs. These examples represent the extreme scenarios disclosed by the results reported in Table~\ref{tab:lattice-parameter}. A careful inspection of these data confirms a general trend in which a closer interaction between the D:A pair and the solvent molecule appears to correlate with an increased D:A separation: (i) both donor and acceptor molecules are closer to the solvent than to each other, and (ii) shorter donor-acceptor distances lead to larger separations between solvent molecules and both DBBTF and HATCN.

The dihedral angles in HATCN represent another significant structural degree of freedom in the considered OCCs. Thanks to its peculiar shape, this acceptor hosts three N$\equiv$C-C=C-C$\equiv $N angles, named here $\theta_i$ ($i = 1,2,3$, see Figure~\ref{fig:scheme}e and Figures~S1 and S2), quantifying the out-of-plane displacement of the cyano groups relative to the central carbon ring. Our results, reported in Table~\ref{tab:lattice-parameter}, reveal significant, crystal structure- and solvent-dependent variations in these distortions. For OCCs containing a single ACE or THF molecule, the dihedral angles are smaller, with $\theta_1=0.9^{\circ}$ and $\theta_2=1.7^{\circ}$ for D:A:ACE$_1$ and $\theta_1=0.4^{\circ}$ and $\theta_2=0.5^{\circ}$ for D:A:THF$1$, while $\theta_3$ is notably larger (6.6$^{\circ}$ in D:A:ACE$_1$ and 8.0$^{\circ}$ in D:A:THF$_1$, see Table~\ref{tab:lattice-parameter}. On the other hand, the D:A:MeCN cocrystal shows a significantly smaller donor-acceptor distance ($d_{D:A}=3.22$~\AA{}) compared to the other 1:1 solvated OCCs, despite a more pronounced acceptor distortion where all cyano groups are displaced, see $\theta_i$ angles in Table~\ref{tab:lattice-parameter}. This apparent contradiction highlights the complex interplay of forces within the crystal.

In OCCs containing two residual solvent molecules, the competition between steric and electrostatic interactions between the solvent and the D:A pairs slightly alters the picture. Differences among the dihedral angles are less pronounced compared to the crystals with a single solvent molecule in the unit cell (Table~\ref{tab:lattice-parameter}). This is apparent in the presence of THF, where the three dihedral angles adopt different values: $\theta_1=1.4^{\circ}$, $\theta_2=2.4^{\circ}$, and $\theta_3=5.5^{\circ}$. Overall, the structures become more compact along the $\pi$-stacking direction, as indicated by the systematic decrease in the lattice vector $a$ with an increasing number of solvent molecules (Table~\ref{tab:lattice-parameter}). Finally, the non-solvated OCC with 3:2 ratio exhibits similar dihedral angles, varying between 3.2-3.3$^{\circ}$ ($\theta_1$ and $\theta_2$) and 4.9$^{\circ}$ ($\theta_3$). This finding suggests a key role of residual solvent molecules in modifying the structural properties of the cocrystals, particularly their molecular conformation and packing.

\subsection{Vibrational analysis and degree of charge transfer}

\begin{figure}
    \centering
\includegraphics[width=1\textwidth]{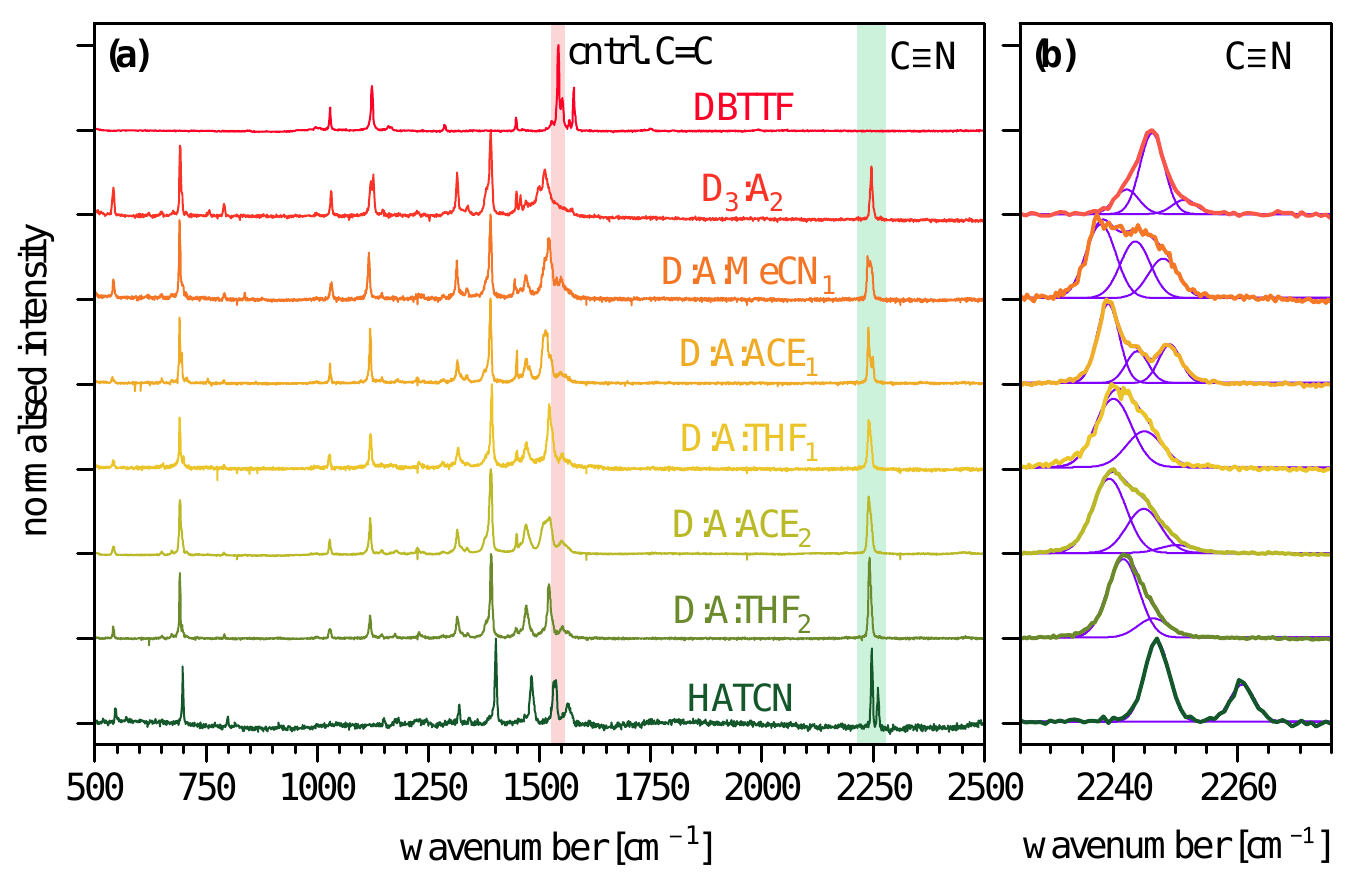}
    \caption{(a) Normalized Raman spectra of all D:A cocrystals and their pristine constituents DBTTF and HATCN. The resonances associated with the central (cntrl.) C=C stretching mode in DBTTF and the C$\equiv$N vibration in HATCN are marked by shaded red and green areas, respectively. (b) Enlarged view of the spectral region of the C$\equiv$N mode, with fitted Gaussian peaks in purple.}
    \label{fig:Raman}
\end{figure}

In the next step of our analysis, we evaluate the DCT of the considered OCCs using Raman spectroscopy~\cite{chap81jacs}. As shown in Figure~\ref{fig:Raman}a, pristine DBTTF is characterized by unique spectroscopic signatures within the 1000-1600~cm$^{-1}$ range, with the most intense resonance at 1543~cm$^{-1}$ assigned to the stretching of its central C=C bond~\cite{Girlando1983,Tanaka1986}.  For HATCN, distinct features are observed at approximately 700, 1300, 1500, and 2250~cm$^{-1}$. The highest-frequency resonance is assigned to the cyano (C$\equiv$N) stretching mode~\cite{Kona+20CEJ}.

In the Raman spectra of the DBTTF:HATCN OCCs, most of the spectral features of the constituents are observable with similar relative intensity and negligible energy shift compared to their isolated form (Figure~\ref{fig:Raman}a). However, the central C=C stretching mode in DBTTF ($\nu=1543 \, \text{cm}^{-1}$, red-shaded area in Figure~\ref{fig:Raman}a) is an exception. This resonance, which overlaps with the carbon-carbon vibrational modes of HATCN, is suppressed in all cocrystals, including those with 3:2 molar ratio. Likewise, no vibrational signatures from the solvent molecules are visible in the Raman spectra, in agreement with previous findings on THF-loaded cocrystals of perylene as donor and 1,2,4,5-tetracyanobenzene as acceptor~\cite{Zhai+21NC}. These combined results suggest the generally low Raman sensitivity of the considered solvent molecules. 

The cyano mode in HATCN at 2250~cm$^{-1}$ appears as a well-defined resonance in the Raman spectra of all considered OCCs (Figure~\ref{fig:Raman}a, green shaded area). Since this feature dominates the high-frequency spectral region without interference or overlap with the carbon modes, it provides a well-suited fingerprint for further analysis of the various DBTTF:HATCN OCCs considered in this work. The spectrum of the HATCN powder exhibits two well-separated peaks assigned to symmetric and antisymmetric stretchings of neighboring cyano groups (Figure~\ref{fig:Raman}b, bottom). With the aid of a Gaussian fit, we identify two or three distinct peaks contributing to the resonance, depending on the specific molar ratio and residual solvent. Specifically, two features appear in the THF-containing cocrystals, while three modes contribute in the presence of ACE and MeCN. This observation, which can be rationalized by the larger dipole moments of ACE and MeCN compared to THF (Table~S6), points to a stronger solvent-D:A interaction in the former two cases. Three Gaussian peaks also contribute to the C$\equiv$N resonance in the spectrum of the D$_3$:A$_2$ polymorph (Figure~\ref{fig:Raman}b, top). In this cocrystal, we interpret the splitting as a consequence of the larger unit cell (Table~\ref{tab:lattice-parameter}) and of the lifted degeneracy of the vibrational energies of the cyano mode, due to the different orientation of the two HATCN molecules in the cocrystal structure (Figure~\ref{fig:scheme}b). 

\begin{figure}
    \centering
\includegraphics[width=0.5\textwidth]{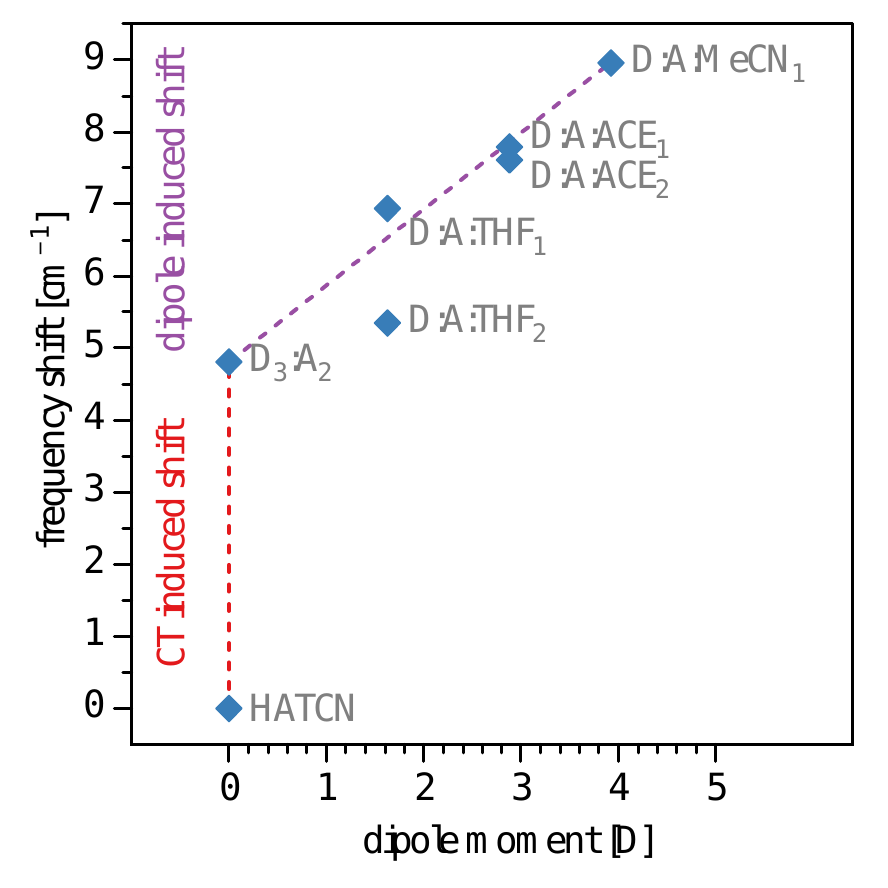}
    \caption{Raman shifts of the lowest-energy C$\equiv$N-mode in the investigated DBTTF:HATCN OCCs, as a function of the dipole moment of the solvent molecules~\cite{NIST_DM}, offset to the value obtained for pristine HATCN. The dashed red line points to the DCT-induced shift, while the violet one indicates the dependence of the solvent polarity.}
    \label{fig:Shift}
\end{figure}

The analysis of the frequency shift of the C$\equiv$N resonance from pristine HATCN to the cocrystals (Figure~\ref{fig:Raman}b) represents an effective means to evaluate the DCT. As shown in Figure~\ref{fig:Shift}, this shift, taken for the low-energy peaks, ranges from 4.8~cm$^{-1}$ in the D$_3$:A$_2$ polymorph to 9.0~cm$^{-1}$ in D:A:MeCN. It stands out that the frequency shift increases with the dipole moment of the incorporated solvent molecules for the cocrystals with D:A:sol$_1$ stoichiometry. The OCCs with two residual solvent molecules in the unit cell depart from this trend, as observed for both ACE, where the shifts in D:A:ACE$_1$ and D:A:ACE$_2$ are very similar, and for THF, where the variation is more pronounced. We attribute this behavior to neighboring THF molecules in the cocrystal being oriented in opposite directions, effectively cancelling the total dipole moment acting on the D:A stack. On the other hand, due to the larger dipole moment of the ACE molecules compared to the THF molecules, this canceling effect appears absent in the cocrystals hosting this solvent, leading to different interaction schemes between D and A molecules.

The vibrational shifts associated with the C$\equiv$N resonance as a function of the dipole moment of the solvents reveal two crucial effects. By setting to zero the reference value of the cyano mode in Figure~\ref{fig:Shift}, we can decouple the shift $\Delta \nu$ into two parts,
\begin{equation}
\Delta \nu = \Delta\nu_\mathrm{DCT}+\Delta\nu_\mathrm{dipole},
\label{eq:delta_nu}
\end{equation} 
one associated with the DCT when in the presence of the donor, and the other one stemming from solvent polarity and increasing linearly with the associated dipole moment $p$. Inserting in Eq.~\eqref{eq:delta_nu} parameters extracted from Figure~\ref{fig:Shift} leads to
\begin{equation}
    \Delta \nu = 4.81~\mathrm{cm}^{-1} +  1.06\,\frac{\mathrm{cm}^{-1}}{\mathrm{Debye}}~p,
\end{equation}
where the contribution due to D:A interaction is $4.81~\mathrm{cm}^{-1}$, while $1.06~\mathrm{cm}^{-1}$ per Debye is given by the slope of the line connecting the results for D$_3$:A$_2$ and D:A:MeCN. 
A linear relation between frequency shift and DCT is well-established in the literature~\cite{Chap+81JACS,Kona+20CEJ}. As a reference, the cyano mode in the anion present in a crystal violet salt is shifted by approximately 45\,cm$^{-1}$ compared to the cyano mode in neutral HATCN~\cite{Kona+20CEJ}. 
Based on our experimental evidence, the DCT between DBTTF and HATCN is estimated to be about $0.11~e$, independent of the polymorph and the solvent used for crystallization. It is worth noting that all values reported in Figure~\ref{fig:Shift} correspond to an upper limit, as the interaction between molecular vibrations and excess electrons in the samples is expected to overestimate the frequency shift and the related DCT~\cite{Girl+07MCLC,Goet+17AEM}.

\begin{table} 
\centering
\caption{Degree of charge transfer in electrons for the investigated cocrystals. The absolute values (asb.) for the computationally simulated solvent-free systems (w/o Sol) are provided for comparison with the values computed for the solvated OCCs, where a positive sign indicates electron depletion and a negative sign indicates electron accumulation on a given molecule.}
\label{tab:DCT}
\begin{tabular}{l|c|cccc}
\hline \hline
Cocrystal & \multicolumn{5}{c}{Degree of charge transfer [$e$]} \\ \cline{2-6}
& w/o Sol (abs.) & DBTTF & HATCN & Sol1 & Sol2 \\ \hline \hline
D$_3$:A$_2$ & 0.13 & -- & -- & -- & -- \\ \hline
D:A:THF$_1$ & 0.13 & 0.17 & -0.01 & -0.16 & -- \\
D:A:THF$_2$ & 0.16 & 0.26 & 0.09 & -0.16 & -0.19 \\ \hline
D:A:ACE$_1$ & 0.13 & 0.15 & -0.07 & -0.08 & -- \\
D:A:ACE$_2$ & 0.14 & 0.17 & 0.03 & -0.10 & -0.10 \\ \hline
D:A:MeCN$_1$ & 0.09 & 0.09 & -0.06 & -0.03 & -- \\
\hline \hline
\end{tabular}
\end{table}

Partial charge analysis from DFT, performed in the Hirshfeld scheme~\cite{hirs77tca}, reveals a charge transfer of 0.13~$e$ in the solvent-free D$_3$:A$_2$ OCC, in very good agreement with the experimental findings discussed above. However, for the solvated polymorphs, a more faceted scenario appears. We start by examining the values obtained by computationally removing the solvent molecules cocrystal structure without performing any further relaxation. In this case, we find similar values around 0.1~$e$ for all systems, in excellent agreement with the experimental estimate from Raman spectroscopy. In these fictitiously binary systems, the negative charge accumulated on HATCN is identical in magnitude but reversed in sign from the one donated by DBTTF. Corresponding absolute values are reported in Table~\ref{tab:DCT}. 

By extending our analysis to the solvated OCCs, in which the solvent molecules are included in the unit cell, we find less trivial trends. Given the ternary composition of these systems, we report in Table~\ref{tab:DCT} the values computed for each unit with the corresponding sign (positive: electron depletion; negative: electron accumulation). While the partial charge on DBTTF is substantially unaffected by the presence of the solvent, except for D:A:THF$_2$, where it donates 0.26~$e$, the results reported for the acceptor are unexpected. In both considered OCCs hosting two solvent molecules, D:A:THF$_2$ and D:A:ACE$_2$, HATCN acts as a donor by releasing 0.09 and 0.03~$e$, respectively (Table~\ref{tab:DCT}). In the presence of a single solvent molecule in the unit cell, the reported partial charge on HATCN is consistently negative, but its magnitude is lower than 0.1~$e$ in all cases. On the other hand, the solvents host large negative charges in all configurations. In D:A:THF$_1$, THF withdraws 0.16~$e$ while in D:A:THF$_2$ the two solvents take up -0.35~$e$ in total. In the latter case, not only DBTTF but also HATCN donate charge to the solvent. A similar scenario, although slightly weaker in magnitude, is obtained with ACE. In D:A:ACE$_1$, the solvent and HATCN withdraw almost the same amount of charge to the donor, which releases 0.15~$e$. In D:A:ACE$_2$, the two solvents combined attract -0.20~$e$ mostly from both DBTTF (0.17~$e$) but also, to a lesser extent (0.03$e$) from HATCN. In D:A:MeCN$_1$, the expected scenario is partially restored with only 0.03~$e$ uptaken by the solvent, representing, however, one third of the total charge donated by DBTTF (Table~\ref{tab:DCT}).

The origin of this non-trivial behavior can be explained by reconsidering the structural analysis of the OCCs and, in particular, the relatively short distances between the solvent species and both DBTTF and HATCN, all below 3~\AA{} (Table~\ref{tab:lattice-parameter}). As illustrated in Figures~S1 and S2, the larger solvent molecules THF and ACE intercalate between the donor and the acceptor, exhibiting the shortest distances to the D:A pair. The N-termination of HATCN plays a crucial role in this process, attracting polar solvents and giving rise to a subsystem in which the solvent becomes the primary electron acceptor. When two THF and ACE molecules are included in the unit cell, this effect is magnified with both DBTTF and HATCN donating charge to the solvent. The smaller size of MeCN compared to THF and ACE (Figure~S1) reduces the strength of this interaction and, consequently, the amount of electronic charge transferred to the solvent. We stress that the physical scenario depicted by DFT points to an excess of charge accumulated on the solvent molecule in the cocrystal rather than a state-selective charge-transfer, as confirmed by the following analysis on the electronic and optical properties of the OCCs. At the same time, it emphasizes that in these ternary crystals, charge distribution is dictated not only by conventional channels such as the cyano groups in HATCN scrutinized by Raman spectroscopy, but also involves the entire atomic network.

\subsection{Optical Absorption and Electronic Structure}

\begin{figure}
    \centering
\includegraphics[width=\textwidth]{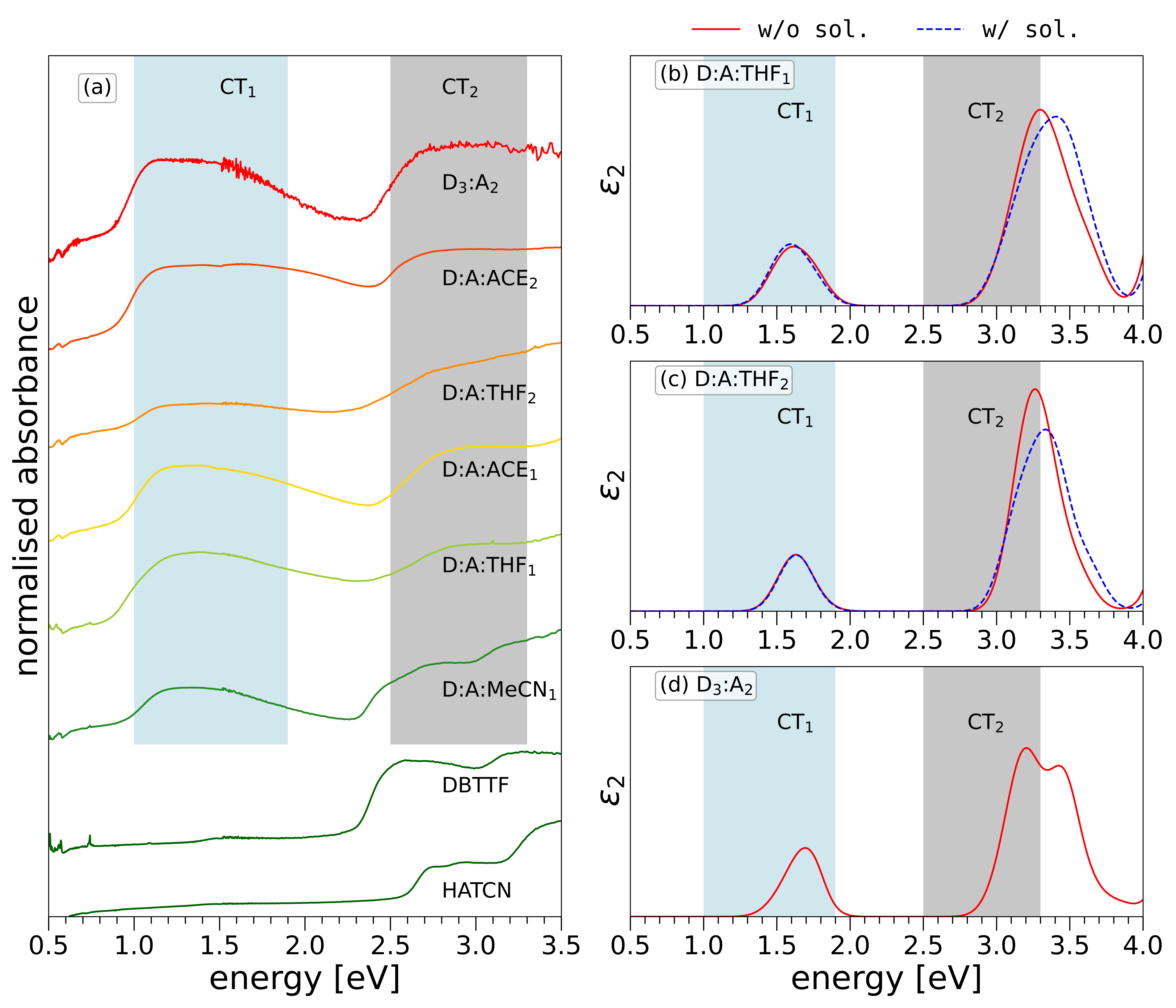}
    \caption{a) Normalized absorbance measured for all D:A cocrystals and for the pristine materials DBTTF and HATCN. Imaginary part of the macroscopic dielectric constant calculated for b) D:A:THF$_1$, (c) D:A:THF$_2$, and (d) D$_3$:A$_2$ OCCs and visualized with a Gaussian broadening of 20 meV. Results obtained with and without the inclusion of solvent molecules are indicated by solid red and dashed blue curves. Light-blue and gray areas highlight the CT$_1$ and CT$_2$ absorption bands, respectively, in both the experimental and computed spectra.}
    \label{fig:Abs}
\end{figure}

A unique feature of D:A OCCs is the presence of low-energy absorption bands that are not present in the spectra of the individual constituents. Usually, they stem from the transition between the HOMO of the donor to the LUMO of the acceptor or, in case of very strong coupling, between hybridized frontier orbitals with bonding and anti-bonding character~\cite{Goris2005,mend+13acie,vale-cocc19jpcc}. The cocrystals investigated in this work are no exception, as shown in Figure~\ref{fig:Abs}a, where their absorbance is displayed with that of DBTTF and HATCN powders. The absorption onset is at 2.28~eV for DBTTF and 2.57~eV for HATCN. In the spectrum of the donor, the lowest absorption band in the visible region is accompanied by another one of similar intensity in the near-ultraviolet (UV), between 3.2 and 3.5~eV. Similarly, the visible absorption band in the acceptor is followed by a stronger feature in the UV, extending beyond 3.5~eV. We note in passing that these spectra contain spurious contributions from background scattering that could not be eliminated even using an integrating sphere to collect diffuse scattered light from powder samples. 

Moving now to the spectra of the cocrystals, we notice that all samples with a 1:1 molar ratio show pronounced absorption between 0.85~eV and 2.25~eV. This feature, which finds no counterpart in the spectra of the constituents (Figure~\ref{fig:Abs}a), can be interpreted as a charge-transfer (CT) excitation, labeled CT$_1$. A second absorption band (CT$_2$) is present between 2.4~eV and 3.3~eV. It is well pronounced in the solvent-free polymorph D$_3$:A$_2$ and the polymorphs with one and two ACE molecules and one residual THF. Due to a strong scattering background, the CT band is smeared out for the D:A:THF$_2$ polymorph. In contrast, the measurement of the D:A:MeCN$_1$ polymorph in this energy region reveals distinct peaks (see also Figure~S6) that can be associated with the absorption signatures of the pristine materials (Figure~\ref{fig:Abs}a). This behavior can be attributed to the fact that DBTTF has the lowest solubility in MeCN among the three solvents used. This triggers the crystallization of pristine DBTTF, pristine HATCN, and the D:A:MeCN$_1$ cocrystal phase in their simultaneous presence in the analyzed powder. 

\begin{figure}
    \centering
\includegraphics[width=0.5\textwidth]{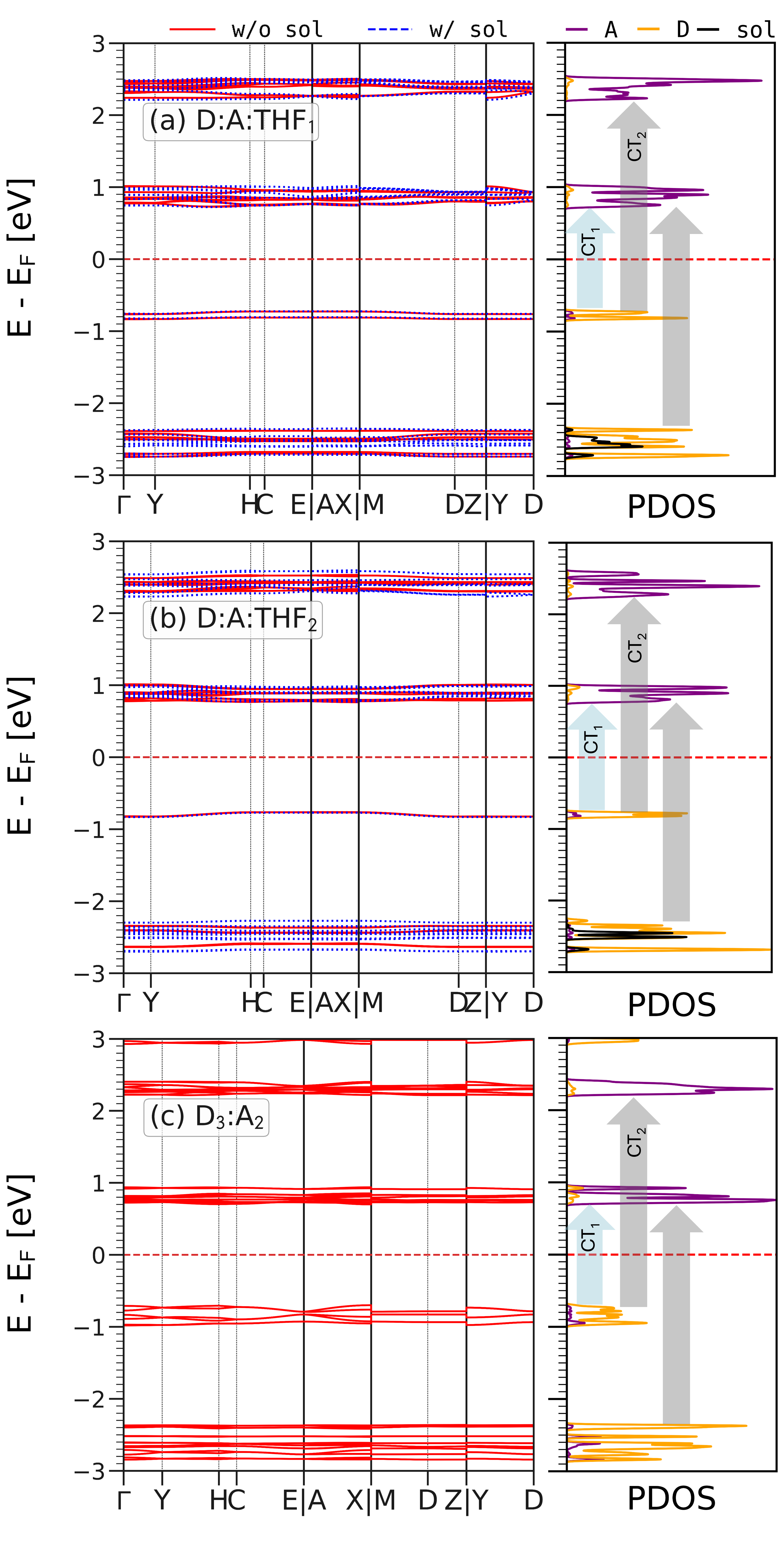}
    \caption{Electronic band structure (left) and PDOS (right) of the (a) D:A:THF$_1$, (b) D:A:THF$_2$, and (c) D$_3$:A$_2$ cocrystals. Results obtained for comparison by removing the solvent molecules from the unit cells of the 1:1 OCCs are shown by dashed blue lines in panels (a) and (b). In the PDOS, contributions from states localized on the donor and acceptor molecules are shown in orange and purple, respectively. The Fermi energy (E$_F$) is set to zero in the mid-gap. Light-blue and gray arrows indicate the transitions giving rise to CT$_1$ and CT$_2$, respectively.}
    \label{fig:BANDs}
\end{figure}

\begin{table} 
\centering
\caption{Energy gap ($E_\mathrm{gap}$) and  the band gap without solvent (w/o sol) are estimated by simulating the OCCs with the solvent molecules removed from the unit cells.}
\label{tab:Egap}
\begin{tabular}{lcc}
\hline \hline
Cocrystal & $E_\mathrm{gap}$ (w/ sol) [eV] & $E_\mathrm{gap}$ (w/o sol) [eV] \\ \hline \hline
D$_3$:A$_2$  & -- & 1.39 \\ \hline
D:A:THF$_1$  & 1.43 & 1.44 \\
D:A:THF$_2$  & 1.53 & 1.52 \\ \hline
D:A:ACE$_1$  & 1.47 & 1.44 \\
D:A:ACE$_2$  & 1.51 & 1.51 \\ \hline
D:A:MeCN$_1$ & 1.48 & 1.46 \\
\hline \hline
\end{tabular}
\end{table}

To better resolve the CT bands in the spectra of the OCCs, we performed \textit{ab initio} calculations in the random-phase approximation on top of DFT (PBE0 functional) of the D:A:THF$_1$, D:A:THF$_2$, and D$_3$:A$_2$ cocrystals with and without solvent molecules.
As shown in Figure~\ref{fig:Abs}b-d, all spectra exhibit absorption peaks in the regions associated with CT$_1$ and CT$_2$ in the experiment. Notably, the energy of CT$_1$ is almost identical among the considered OCCs, consistently appearing in the range 1.40 - 1.55~eV, irrespective of the molar ratio, number of solvents, or cocrystal structure. The second absorption peak, associated with CT$_2$, is slightly blue-shifted compared to the measurements, an expected consequence of the adopted computational scheme~\cite{cocc-drax15prb}, which does not explicitly account for electron-hole attraction. In this case, the influence of solvents is more pronounced, affecting both energy and intensity (Figure~\ref{fig:Abs}b-d). In the spectra of the 1:1 cocrystals, the peak is centered at approximately 3.5 and 3.4~eV, in the presence of one and two solvent molecules, respectively (Figure~\ref{fig:Abs}b,c). In an idealized structure without solvent, CT$_2$ red-shifts by about 100~meV in both spectra, slightly gaining in relative intensity for D:A:THF$_2$. In the spectrum of the 3:2 cocrystal, the CT$_2$ band is composed of two distinct peaks (Figure~\ref{fig:Abs}d).

To gain deeper insight into the optical characteristics of these OCCs, including the origin of the absorption peaks, we inspect their electronic band structure and projected density of states (PDOS) computed from DFT. The results displayed in Figure~\ref{fig:BANDs} indicate a staggered alignment in all three systems. The highest-occupied and lowest-unoccupied levels are localized on the donor and acceptor molecules, respectively. A similar trend is found for the other OCCs, see Figures~S3 and S4. The minimally dispersive character of these bands confirms the molecular nature of the electronic states even in the cocrystalline form. The inclusion of the solvent in the 1:1 stoichiometry plays no role at the frontier, since contributions from THF appear only deeper in the valence region (Figure~\ref{fig:BANDs}a,b; Figures~S3 and S4 for the other solvents). By comparing the excitation energies of the absorption bands in Figure~\ref{fig:Abs}b-d, we can assign each feature to specific manifolds of electronic transitions. As indicated by the light-blue arrow, in all OCCs, CT$_1$ stems from the excitation between the highest-occupied bands localized on DBTTF and the lowest unoccupied ones residing on HATCN molecules (Figure~\ref{fig:BANDs}). On the other hand, CT$_2$ is generated by two transitions involving the highest occupied band and the second-lowest unoccupied one, as well as the second-highest valence band to the lowest unoccupied one (gray arrows in Figure~\ref{fig:BANDs}). It is worth noting that one of the CT$_2$ excitations starts from a valence band manifold hosting contributions from the solvent. Some of these states are entirely localized on the solvents, while others are also distributed on the donor (see Figure~S5). 

\begin{figure}
    \centering
\includegraphics[width=1.0\textwidth]{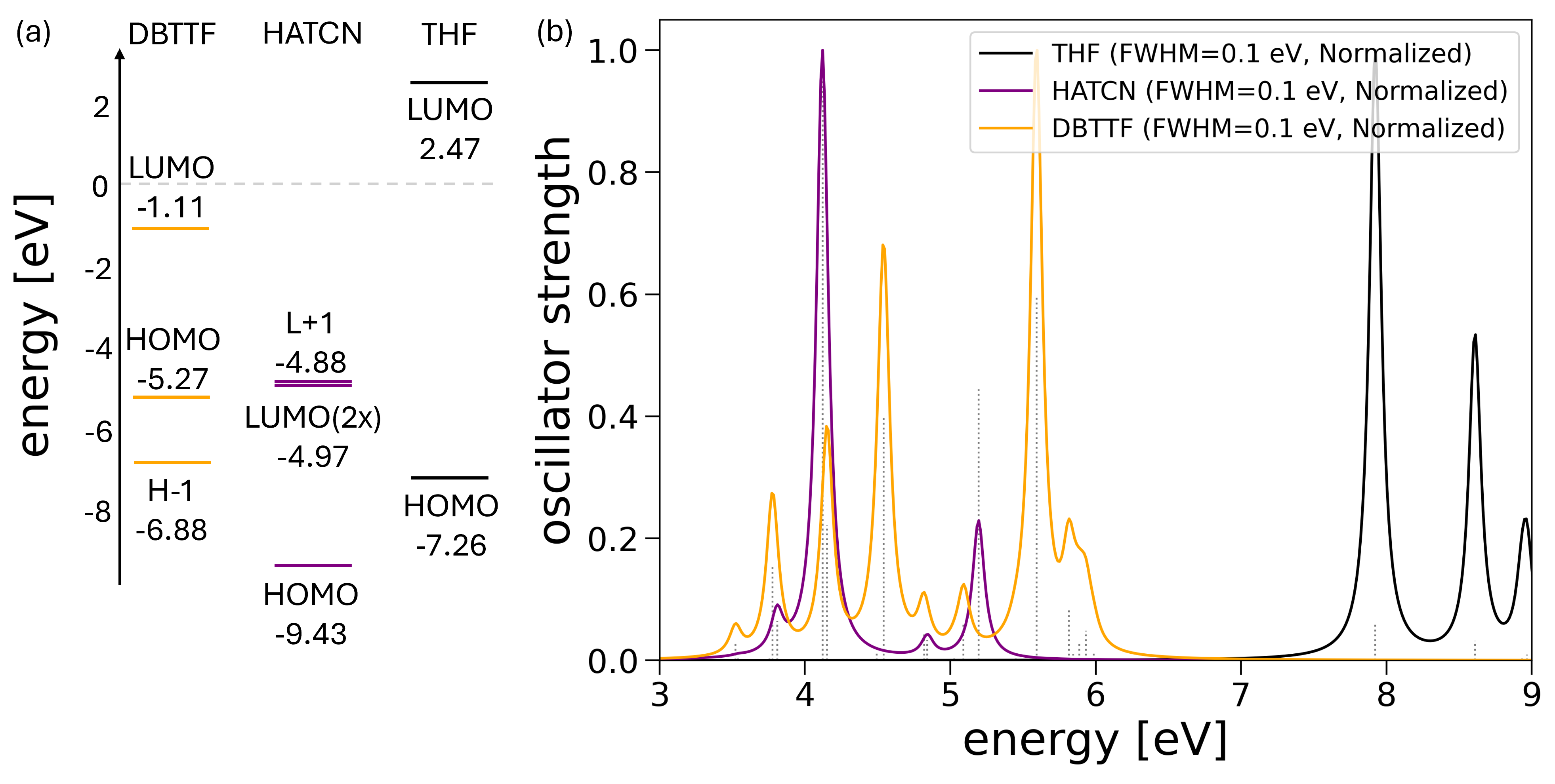}
    \caption{a) Energy levels with respect to the vacuum level set to zero (gray dashed line) of DBTTF, HATCN, and THF in the gas phase from DFT (PBE0 functional). b) Normalized Lorentzian-broadened optical absorption spectra (full-width half-maximum of 0.1~eV) of DBTTF, HATCN, and THF. The $\delta$-like (unsmeared) electronic transitions are marked by gray dashed bars.}
    \label{fig:isolated-molecules}
\end{figure}

The character of the electronic states highlighted in the PDOS (Figure~\ref{fig:BANDs}) is consistent with the relative energies of the highest (lowest) occupied (unoccupied) levels in the isolated molecules.
As illustrated in Figure~\ref{fig:isolated-molecules}a, both the HOMO and the HOMO-1 of DBTTF are within the HOMO-LUMO gap of HATCN. Interestingly, the HOMO of THF is below but not far from the HOMO-1 of the donor. The HOMO-LUMO gap of the solvent is close to 10~eV and, as such, substantially larger than those of the donor and acceptor molecule, both on the order of 4~eV. These electronic characteristics are reflected in the optical absorption spectra of the gas-phase molecules, shown in Figure~\ref{fig:isolated-molecules}b. While the results obtained for DBTTF and HATCN largely overlap in the near-UV region, the absorption peaks of THF appear further in the UV, starting from approximately 8~eV. 

The most striking difference between the electronic structure of the OCC with 1:1 and 3:2 ratios is exemplified in Figure~\ref{fig:BANDs}. The highest occupied levels in both D:A:THF$_1$ and D:A:THF$_2$ are flat bands, with single-molecule character (Figure~\ref{fig:BANDs}a,b). In contrast, in the D$_3$:A$_2$ cocrystal, the uppermost valence states form a six-state manifold with a bandwidth of 280~meV. This is a direct result of the molecular arrangement, promoting hybridization between the HOMOs of the donor molecules and the LUMOs of acceptors. 

This analysis leads us to further considerations regarding the size of the fundamental gaps in the OCCs and the dependence of this quantity on the crystal structure and type and amount of solvent molecules in the unit cell. The results listed in Table~\ref{tab:Egap} point to minimal variations among the considered cocrystals with the 1:1 molar ratio, all exhibiting band gaps fluctuating around 1.5~eV, while the 3:2 OCC has a slightly smaller band gap of 1.4~eV. Increasing the number of solvent molecules leads to consistently larger band gaps with differences on the order of 0.1~eV. We interpret this finding in terms of the structural variations in the lattice and the consequently modified intermolecular interactions caused by the residual solvents. As summarized in Table~\ref{tab:lattice-parameter}, the unit-cell volume systematically increases with the number of solvent molecules. A crystalline expansion generally diminishes electronic screening. In these D:A cocrystals, it additionally promotes short-range intermolecular couplings at the expense of long-range electronic interactions, overall leading to a band-gap increase~\cite{cocc+22jpm}.

\section{Summary and Conclusions}
In this joint experimental and computational study, we investigated six DBTTF:HATCN (solvated) cocrystals, revealing the impact of solvent molecules on the degree of charge transfer as well as on the electronic and optical properties. Despite expected variations related to the inclusion of solvent molecules, the fundamental energy gap is largely unaffected by the crystal structure and molar ratio, suggesting that the electronic structure is robust against solvent interference. Likewise, the absorption spectra for all cocrystals exhibited striking similarities. The absorption onset (1.39 - 1.53 eV across all considered OCCs) is dominated by a charge-transfer excitation that is not present in the spectra of the constituents. Another charge-transfer excitation appearing at higher energies between 2.5 and 3.3~eV is more influenced by the amount of solvent molecules, especially THF and ACE, underscoring the subtle yet significant role solvents play in modulating excited-state properties beyond the fundamental energy gap. The electronic transitions responsible for these excitations are identified from theory, revealing the peculiar offset of the electronic levels of the constituents as responsible for the appearance of two charge-transfer excitations in the spectra of the cocrystals. The calculations also reveal the negligible influence of electronic transitions in the solvents, characterized by optical gaps in the far UV. 

A detailed analysis of the Raman spectra and particularly of the C$\equiv$N mode in HATCN as a function of the solvent polarity enables evaluating the degree of charge transfer in the examined cocrystals. The value of 0.11~$e$ extracted from the measurements is in excellent agreement with the one obtained from partial-charge analysis of the DFT results if the role of solvent is neglected. However, simulations of the ternary cocrystals reveal a more complex and unexpected role for the solvent molecules. The most striking finding of this work is the direct electronic participation of the solvent in the charge-transfer process within the OCCs, with the solvent molecules acting as the primary electron acceptors. Their ability to acquire a large negative charge is so pronounced that it affects donor-acceptor interactions. While this result may seem contradictory in relation to the experimental evidence, it actually reveals a subtle yet crucial mechanism, such as the participation of all atoms in the charge-transfer process, not just the cyano group of HATCN, as probed by Raman spectroscopy. From a methodological perspective, this finding reinforces the importance of complementary experimental and theoretical analysis to disclose and interpret the fundamental properties of these novel, complex materials.

In conclusion, this study provides an in-depth understanding of solvent-intercalated DBTTF:HATCN cocrystal polymorphs. While DBTTF:HATCN crystallization via technologically relevant solution-based methods does not deteriorate the inherent characteristics of these cocrystals, the amount and type of solvent included in the unit cells can be leveraged as an effective and non-invasive knob to tune the molecular arrangement and, hence, the intrinsic properties. This discovery highlights a new paradigm for designing next-generation organic cocrystals, where solvents are not merely passive structural templates but are strategically employed as active components to optimize the performance of the material in optoelectronic devices.

\section{Experimental Methods}

\subsection{Materials and Crystal Preparation}
DBTTF was purchased from Sigma-Aldrich Chemie GmbH, and HATCN was purchased from Ossila Ltd. The powders were used without further purification for cocrystal growth by solvent evaporation. The solvents THF, ACE, and MeCN were purchased from Carl Roth GmbH + Co. KG with purities higher than 99.9~\%.
The pristine molecular powders were dissolved in the respective solvents at elevated temperature (hot plate 60 $^{\circ}$C) and stirred. It was observed that the solubility of HATCN is good in all three solvents. In contrast, the solubility of DBTTF is good only in THF and is decreased in ACE and even lower in MeCN. After completely dissolving the solution, it was stored to reach room temperature. After mixing the solutions, solvent evaporation was allowed at room temperature to grow crystals. The low solubility of DBTTF in ACE and MeCN results in the precipitation of pristine DBTTF first and cocrystals later. As the cocrystals appear dark green in all cases, the separation from slightly orange DBTTF crystals and yellow HATCN crystals from the cocrystal phase is directly possible. Precipitation of pristine materials was absent for cocrystal growth from equimolar DBTTF:HATCN mixtures dissolved in THF and ACE. Drying at elevated temperatures was performed by placing the solution in a test tube immersed in glycerine, heated on a hot plate. The hot plate temperature of 140$^{\circ}$C results in 80$^{\circ}$C of the glycerine, which was stirred the whole time. This preparation was performed for THF and ACE as solvents to reduce the number of solvent molecules inside the crystal. All powders of the cocrystals contain green to dark-green needle-like crystallites. 

The method of horizontal vapor deposition \cite{laud+98jcg} is employed to obtain solvent-free \linebreak DBTTF$_3$:HATCN$_2$ cocrystals under a laminar nitrogen gas flow of 30 sccm. HATCN powder was used as received, while DBTTF was additionally purified via gradient sublimation before crystal growth. To enable co-evaporation despite the substantial difference in sublimation temperatures of HATCN (680~K) and DBTTF (490~K), the source materials were placed spatially separated inside the furnace at positions with appropriate temperatures, which were determined beforehand. Applying a steep temperature gradient in the crystal growth zone facilitated a spatially defined recrystallization region, leading to the growth of dendritic cocrystals of green color with dimensions of up to 2~mm in length and several hundred microns in width. Alongside the cocrystals, pure HATCN and DBTTF recrystallize at higher and lower temperature regions.

\subsection{Crystal Structure Determination}

\subsubsection{Details about Crystals Grown by Horizontal Vapor Deposition}

A D$_3$:A$_2$ crystal suitable for single-crystal X-ray diffraction was selected, coated in perfluoropolyether oil, and mounted on a polyimide microloop. Diffraction data were collected on a RIGAKU XTALAB SYNERGY-R diffractometer with an HPA area detector (HyPix-Arc150) using multi-layer mirror monochromated Cu K$\alpha$ radiation generated by a rotating-anode PhotonJet-R X-ray source. The crystal was cooled using an Oxford Cryostreams low-temperature device. Data was collected at 100 K. The images were processed and corrected for Lorentz-polarization effects and absorption as implemented in the CrysAlis$^{Pro}$ software of the Rigaku company. The structure was solved using the intrinsic phasing method (SHELXT) \cite{shel15acsa} and the Fourier expansion technique. All non-hydrogen atoms were refined in anisotropic approximation, with hydrogen atoms 'riding’ in idealized positions, by full-matrix least-squares against $F^2$ of all data, using SHELXL\cite{shel15acsc} software and the SHELXLE graphical user interface\cite{Hubschle:kk5092}. 

\subsubsection{Details about Crystals Grown by Solvent Evaporation}

The data collections were performed with a BRUKER D8 VENTURE diffractometer with an area detector using Mo K$\alpha$ radiation. Multi-scan absorption corrections implemented in SADABS \cite{SADABS} were applied to the data. The structures were solved by the intrinsic phasing method (SHELXT-2013 \cite{shel15acsa}) and refined by full-matrix least-squares procedures based on F$^2$ with all measured reflections (SHELXT-2014 \cite{shel15acsc}) in the graphical user interface SHELXle \cite{Hubschle:kk5092} with anisotropic temperature factors for all non-hydrogen atoms. All hydrogen atoms were added geometrically and refined by using a riding model. 

\subsubsection{Crystallographic Information}
Crystal data and experimental details regarding growth are listed in Tables S2-S5; full structural information has been deposited with the Cambridge Crystallographic Data Centre. CCDC number 2473715 and CCDC numbers 2475667--2475671. These data can be obtained free of charge from The Cambridge Crystallographic Data Centre via \url{www.ccdc.cam.ac.uk/data_request/cif}.

\subsection{Raman and Optical Measurements}

Raman spectroscopy was performed on a microphotoluminescence setup, and an excitation wavelength of 532 nm was used to measure the powders using a Horiba LabRAM HR800 setup.

Spectra of optical absorbance were obtained with a Lambda 950 UV/vis/NIR spectrophotometer (Perkin Elmer Inc.) using a 100 mm integration sphere to collect the total transmission, containing direct transmission and diffuse scattered light, for the powder samples. The powders were placed on a microscope slide and pressed with another microscope slide to ensure their adhesion to the slides. Afterward, the slides were separated, and one slide was placed on the spectrophotometer to be measured.

\section{Computational Details}
All \textit{ab initio} calculations presented in this work were performed with the all-electron code FHI-aims~\cite{blum+09cpc}, implementing numeric atom-centred orbital basis sets~\cite{havu+jcp09}. The ``tight'' settings are used along with a 6$\times$5$\times$3 \textbf{k}-mesh to sample the Brillouin zone of each polymorph, according to the conventional paths proposed in Ref.~\citenum{sety-curt10cms}. 
The cocrystals are simulated with  experimental lattice parameters and atomic coordinates. Since hydrogen atoms are not resolved in the measurements, they are included in the input of calculations at standard distances. The resulting structures are relaxed by keeping the volume fixed and constraining all other atomic positions until the residual interatomic forces are smaller than 0.02~eV/\AA{}. The PBE functional~\cite{perd+96prl} is used during optimization, while PBE0~\cite{perd+96jcp} is adopted to evaluate the electronic structure, overcoming the well-known issue of PBE in predicting band gaps. The in-house developed Python library \texttt{aim$^2$dat} \cite{aim2dat} is used to calculate the minimum separation between donor and acceptor complexes across all structures, following the approach successfully adopted in previous work~\cite{edza+23jpcc,sass+24ic}. Partial charge analysis is performed with the Hirshfeld scheme~\cite{hirs77tca} as implemented in FHI-aims. Optical absorption spectra are computed in the random-phase approximation using a Gaussian broadening of 0.02~eV.

\section{Data availability statement}
All data produced in this work are available free of charge at \url{https://doi.org/10.5281/zenodo.15853004}.

\section{Notes}
The authors declare no competing financial interest.

\begin{acknowledgement}
A.M.V. thanks Joshua Edzards for the helpful discussions. A.M.V. and C.C. acknowledge financial support from the German Federal Ministry of Education and Research (Professorinnenprogramm III) as well as from the State of Lower Saxony (Professorinnen f\"ur Niedersachsen). Computational resources were provided by the North-German Supercomputing Alliance (HLRN), project bep00132, and by the high-performance computing cluster CARL at the University of Oldenburg, funded by the German Research Foundation (Project No. INST 184/157-1 FUGG) and by the Ministry of Science and Culture of Lower Saxony. A.O. appreciates funding of project 239543752 by the German Research Foundation. J.P. acknowledges financial support from the Bavarian State Ministry for Science and the Arts within the collaborative research network “Solar Technologies go Hybrid (SolTech)”. L. S.-M. appreciates support from the W\"urzburg-Dresden Cluster of Excellence on Complexity and Topology in Quantum Matter, ct.qmat (EXC 2147). M.S. gratefully acknowledges financial support from the Deutsche Bundesstiftung Umwelt (DBU). A.F. acknowledges financial support from the Julius-Maximilians-Universität Würzburg.
\end{acknowledgement}

\begin{suppinfo}
In the Supporting Information, we report all structural details of the synthesized cocrystals and additional analysis on their electronic and optical properties computed from first principles.

\end{suppinfo}


\providecommand{\latin}[1]{#1}
\makeatletter
\providecommand{\doi}
  {\begingroup\let\do\@makeother\dospecials
  \catcode`\{=1 \catcode`\}=2 \doi@aux}
\providecommand{\doi@aux}[1]{\endgroup\texttt{#1}}
\makeatother
\providecommand*\mcitethebibliography{\thebibliography}
\csname @ifundefined\endcsname{endmcitethebibliography}
  {\let\endmcitethebibliography\endthebibliography}{}

\end{document}